\documentstyle[11pt,newpasp,twoside,epsf]{article}
\def\edcomment#1{\iffalse\marginpar{\raggedright\sl#1\/}\else\relax\fi}
\reversemarginpar

\begin{document}

\title{XDR Chemistry in the Circumnuclear Disk of  NGC~1068}

 \author{A. Usero, S. Garc\'{\i}a-Burillo, A. Fuente}
\affil{Observatorio Astron\'omico Nacional (OAN), C/ Alfonso XII 3, 28014 Madrid, Spain} 

\author{J. Mart\'{\i}n-Pintado}
\affil{Departamento de Astrof\'{\i}sica Molecular e Infrarroja, IEM, CSIC, C/ Serrano 121, 28006
Madrid, Spain} 

\begin{abstract}
 We have studied the feedback influence that the central engine of the Seyfert
2 galaxy NGC~1068 may have on the chemistry of the 200~pc circumnuclear gas disk (CND).  
With this purpose, we have conducted a multi-species/multi-transition survey 
of molecular gas in the CND of NGC~1068 using the IRAM 30m telescope. 
Abundances of several molecular species have been estimated, including HCN, 
CN, CS, HCO$^+$, SiO and HOC$^+$. We report on the detection of significant SiO 
emission in this galaxy, as well as on {\sl the first extragalactic detection 
of the active radical} HOC$^+$. We conclude that the chemistry of the 
molecular gas reservoir in the CND can be best explained in the framework of 
X-rays Dominated Regions (XDR) models.    
\end{abstract}

 Increasing observational evidence suggests that nuclear winds and X-rays may have a
disruptive  influence on the molecular gas in the vicinity of the supermassive Black Holes of
AGN.
XDR could be the dominant sources of emission
for molecular gas in the CND of active galaxies.
 The first evidence of AGN-driven chemistry was provided by
the large HCN/CO ratio $\sim0.01$ found by  Sternberg, Genzel \& Tacconi (1994) in
the CND of NGC~1068. 
Different explanations have been advanced to account for this anomalous HCN chemistry.
Selective depletion of gas-phase oxygen in the dense molecular clouds would explain the measured
high HCN abundance (Sternberg et al. 1994). 
Alternatively, an increased X-ray ionization of molecular clouds near the AGN could enhance the
abundance of HCN
 (Lepp and Dalgarno 1996).

While the aforementioned scenarios intend to reproduce the measured enhancement
of HCN in NGC~1068, their predictions about the abundances of some molecular species differ
significantly.
In order to
constrain observationally the choice of the optimum scenario, we have obtained new IRAM
30m spectra  towards the CND for some of these molecular species. The list includes SiO, HCO$^+$,
H$^{13}$CO$^+$, HOC$^+$ and CN (Usero et al. 2003). We have run LVG models based on the results of
these observations and on 
previously published data (CO: Schinnerer
et al. 2000; HCN and CS: Tacconi et al. 1994, 1997). We have estimated that abundances of
oxygen-bearing species are similar to those of non-oxygenated ones (e.g.: we measure
HCN/HCO$^+$$\sim$1). These results are at odds with the predictions of oxygen
depletion models. On the contrary, the overall abundances fit the predictions of XDR chemistry
schemes (Lepp \& Dalgarno 1996).

The observed SiO emission implies large abundances of this molecule
(SiO/H$_2$ $\sim$5$\times10^{-9}$-$10^{-8}$) in 
the CND. An enhancement of SiO abundances in XDR is expected if the  intense X-rays are able to
evaporate the small ($\sim$10~\AA) silicate
grains and  subsequently increase the  Si fraction in gas phase.

The observation of HOC$^+$ in NGC~1068 is the first extragalactic detection of this
active radical. From LVG models we infer that HOC$^+$ is highly abundant in the CND of this galaxy: we measure a
HCO$^+$/HOC$^+$ ratio $\sim50$, which is even lower than the smallest values found in galactic PDR
(Fuente et al. 2003).  HCO$^+$/HOC$^+$ ratios  in XDR may be smaller than those measured
in PDR due to the higher ionization degrees that take place in the former regions (Smith et
al. 2002).   

\begin{figure}
\plotfiddle{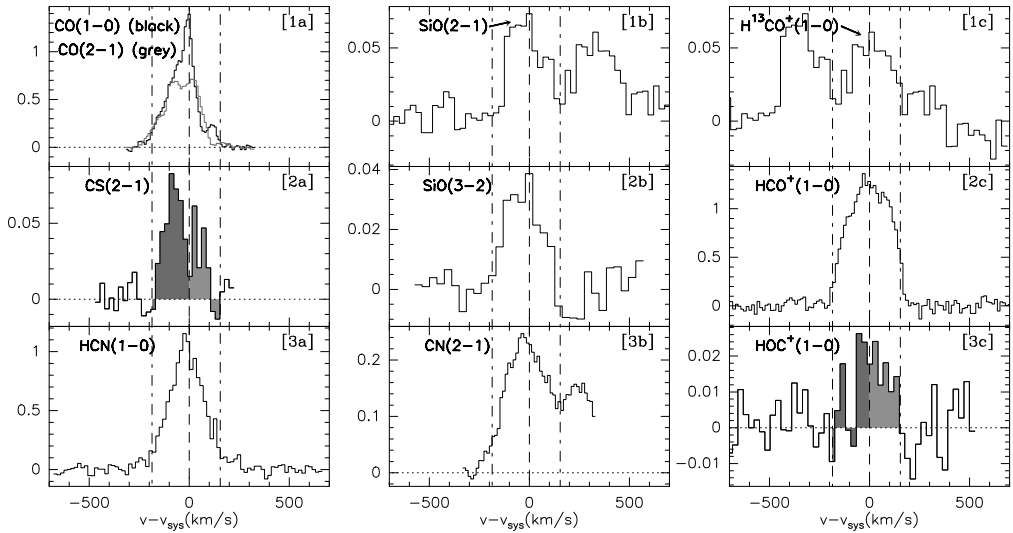}{2.4in}{-90}{57}{57}{-220}{245}
\caption{30m Spectra of the Circumnuclear Disk (CND) in NGC~1068. Y-axis units are T$_{\mathrm{MB}}$
in Kelvin and X-axis units are radial velocities referred to
v$_{\mathrm{sys}}^{\mathrm{LSR}}$=1126km/s.
(from Usero et al. 2003)
} 
\end{figure}

\end{document}